# IMAGERY AGNOSIA
# AND ITS PHENOMENOLOGY

WŁODZISŁAW DUCH
Department of Informatics, Faculty of Physics, Astronomy,
and Informatics, and Neurocognitive Laboratory, Center of Modern
Interdisciplinary Technologies, Nicolaus Copernicus University,
87-100 Toruń, Poland[1]

Lack of vivid sensory imagery has recently become an active subject of research, under the name of aphantasia. Extremely vivid imagery, or hyperphantasia, is at the other end of the spectrum of individual differences. While most research has focused on visual imagery in this paper I argue that from a neuropsychological perspective this phenomenon is much more widespread, and should be categorized as imagery sensory agnosia. After over twenty years of learning to play music phenomenology of auditory imagery agnosia is described from the first-person perspective. Reflections on other forms of imagery agnosia and deficits of autobiographical memories are presented and a hypothesis about putative brain processes that can account for such phenomena is discussed. Extreme individual differences in imagery and in autobiographical memory have implications for many fields of study, from consciousness research to education.

## INDIVIDUAL DIFFERENCES IN IMAGERY

Many research tools to evaluate vividness and clarity of imagery have been developed. First was the Questionnaire on Mental Imagery (QMI) in seven sensory modalities by Betts (1909). Several questionnaires probing auditory imagery have been developed in the past decades (see the review in Halpern, 2015; Hubbard, 2013; Tużnik & Francuz, 2019). The Vividness of Visual Imagery Questionnaire (VVIQ) (Marks 1973;1995) is the most commonly used to estimate the ability to re-create visual experiences. The design of the Bucknell Auditory Imagery Scale (BAIS) (Halpern, 2015) was guided by the construction of VVIQ. The Plymouth Sensory Imagery Questionnaire (Andrade et al., 2014) covers many sensory modalities, including vision, sound, smell, taste, touch, bodily sensation, and emotional feeling. Objective measures of individual variability in vividness of visual mental

---

imagery are based on correlation of early visual cortex activity and the performance on a novel psychophysical task (Cui et al, 2007; Zatorre & Halpern, 2004).

Individual differences in imagery can have a profound effect on learning, understanding of artistic and mathematical talent. Measures of some aspects of mental imagery could provide significant correlates of specific abilities, and may have an important impact on general education. Although research on mental (mostly visual) imagery has a long history, starting with Galton (1880), extreme cases – hyper-vivid or complete lack of imagery – have never been investigated until quite recently. Faw (2009) was the first author who has described non-imagers, basing on his own experience. His reports have met with incredulity. Using questionnaires based on Betts to test 2500 students he has found between 2-5% of people with extremely low imaging scores (Faw, 2009, p. 57). An interesting case of a patient who suddenly lost the ability to generate visual images was described by Zeman et al. (2010), and the term "aphantasia" has been coined to describe people who have "blind imagination" (Zeman, Dewar & Della Sala, 2016). Since that time aphantasia became quite popular, with a quickly growing number of papers focused mainly on neuroimaging of visual imagery (Winlove et al., 2018, Zeman et al. 2020; Milton et al. 2021). Personal aspects of aphantasia and its influence on severely deficient autobiographical memory (Palombo et al., 2015) have been described in a paper by physicist Nicholas Watkins (2018).

I have focused mainly on the lack of auditory imagery, writing about it since 2009 (Duch, 2009; 2011; 2012; 2013). Faw (2009) has analyzed the history of believes that thinking without mental images is not possible. Denial or attempts to explain the phenomenon away led to the lack of interest. Unable to convince psychologists that this subject is worthy of investigation I have been trying to understand my own limitations and experimented with learning to play music. This paper has two goals: in the first part I will describe the phenomenology of auditory perception without imagery, and in the second part I'll argue that aphantasia should be understood as imagery agnosia, with many different subtypes that depend on specific neurological mechanisms. Attempts to create new names, like aphantasia or more recently "dysikonesia" (Dance, Ward & Simner, 2021), do not make much sense from the point of view of neuropsychology and neural mechanisms behind such phenomena.

**PHENOMENOLOGY OF AUDITORY IMAGERY AGNOSIA.**

Faw (2009) has realized that his cognition is not typical when he was 35. I have been over 40 when I started to wonder why my artistic abilities are so poor. In high school, I was fascinated by art and spent countless hours trying to draw. My mother was a professional artist, painting realistic portraits and landscapes. Some of my not-very-bright fellow students could draw very well imaging various scenes, but I have failed even drawing from model objects. I have also tried to learn music, taking lessons on the piano, but after one year had to stop. My musical ear seemed to be fine, I have learned musical notation, but I could not learn to play anything by heart. Yet I loved all sorts of music and knew a lot of song lyrics. I was helping to organize the first discotheques in our school. I have studied physics and made my Ph.D. in theoretical physics, and near the end of the last century, I became interested in cognitive science.

To understand why I have such difficulty with art and music I started playing wind simple instruments, first the recorder (straight flute) and then around 1998 Akai EWI (electronic wind instrument). It can be played in a similar way as the recorder, but can produce many synthetic sounds. I was hoping to improve my musical abilities, treating learning music as a kind of scientific experiment. This hope was based on Aleman et al. (2000) paper: "We hypothesized that music training may be associated with improved auditory imagery ability" (p. 1664). Results of their experiments looked promising: "The musically trained group did not only perform better on the musical imagery task, but also outperformed musically naive subjects on the non-musical auditory imagery task" (p. 1664). On the other hand, attempts to train visual imagery strength was not successful, but the practice has improved metacognitive understanding (awareness of one's own thought processes) related to mental imagery (Rademaker & Pearson, 2012).

I have also found a Doctor of Musical Arts thesis "Mental Representations in Clarinet Performance: Connections Between Auditory Imagery and Motor Behaviors" (Allen, 2007). This work concluded that musicians need to express sounds that they imagine and connect them with a "feel" of motor actions that will produce proper sounds. Good musicians express what they have in mind anticipating what they will hear. When they read music and imagine how it will sound event-related potentials (ERPs) generated by their brains look like those during auditory note perception (Hubbard, 2019).

Reading such advice I gradually started to suspect why I find it so difficult to become a good musician. My rating of vividness on all Bucknell Auditory Imagery Scale items (BAIS, Halpern 2015) is consistently at the "no image" bottom of the scale. These questions do not rise experience of the sound, but a feeling that I know these sounds and will recognize them when I hear them. This "feeling of knowing" has no sensory components. I can try to imitate some sounds if I hear them, for example singing along with the saxophone sounds, but I cannot anticipate the sound that an instrument will make. Without auditory imagery it is impossible to connect motor action with anticipation of the sound that this action will produce, as there is no conscious anticipation. I hear a single sound and try to reproduce it with proper fingering on my instrument. To recognize which sound it was I need to recall its pitch, but if I can't imagine the pitch of the sound I cannot compare it with the one that I produce pressing the keys. All I know is that some keys on the piano (of fingerings on a wind instrument) will produce high or low sounds. I can tune a guitar only if the tuning frequency is played continuously, allowing me to compare it with the sound made by the string.

Recognizing instrument timbre (for example, flute or trumpet sound) is in most cases automatic and effortless. Even in the case of congenital amusia timbre recognition is usually not impaired, indicating that this information is processed in a different way than pitch determination (Marin, Gringas & Stewart, 2012). The timbre of instruments are probably automatically categorized, so there is no need to imagine their sound to recognize them. In the case of amusia different pitch sounds are categorized only in a very broad way: low, middle, or high. In auditory agnosia recognition may be fine (although we do not know yet how good it may be), but there is no sensory experience when the pitch or timbre of a particular instrument is recalled. EWI connected to a MIDI synthesizer is capable of producing many sounds similar to natural wind, brass, string, and other instruments, called "voices". Changing voice I have a feeling of knowing what comes and will feel surprised if the voice I select does not match my anticipation. The strong feeling of knowing what should come, although not being able to imagine it, maybe due to my familiarity with different voices acquired over many years of playing.

One aspect of auditory imagery agnosia is similar to color anomia: people can correctly evaluate the similarity of colors arranging them in rainbow-like patterns, but do not recall the names of colors when they see them. Color naming and color categorization are performed by different brain networks (Siuda-Krzywicka et al., 2021). We do not know how vividly can people with color anomia imagine different colors. If I could recognize and name different pitches of sound I could learn melodies by heart like learning poems, without the need to imagine a sequence of sounds. Categorization without imagery and precise naming of sensory experience is quite poor not only in the case of auditory stimuli. In the case of visual imagery agnosia showing people many objects belonging to the same category and with similar sizes, that differ only by color or small shape variation, may lead to a similar problem with recognition. If objects are significantly different, if they may be assigned to a different category, recognition will be reliable. Sounds that differ only by pitch are hard to categorize, but sounds that differ by timbre belong to different categories. Experiments with change blindness, in which colors or shapes of objects change slowly, show that recognition of gradual changes is rarely noticed (Simons and Rensink, 2005). Some aspects of melodic structures, general rhythmic patterns or melodic contours, may also be remembered, allowing for recognition of melodies. Precise memory for colors and pitches may help to order such percepts, estimate their similarity and assign names, but this is a bottom-up process, information flow from sensory cortices to the associative cortex. To influence vividness or clarity of imagination we need top-down processes, activations flowing from memory to sensory cortices. Visual and auditory imagery agnosia may share some characteristics, but they are different phenomena than color and pitch anomia.

Playing from the musical score is not a problem, transformation of perceived notes to motor functions does not require imagery. I have reached motor proficiency playing my EWI in less than a year and even dared to play J.S. Bach "Erbarme Dich" aria at a conference in Japan (participants were encouraged to play at a formal concert). Repeating melody after hearing it is largely a matter of trial and error. Without the ability to imagine the sound that my action will produce each note is a step into the unknown. The lack of anticipation based on inner hearing gives an impression of no conscious influence on what is being played. As a player I am thus in a similar situation to a listener, frequently surprised myself by what I play. Sometimes a music tune seems to pop up in my head and I feel the need to play, but to hear what goes on in my head I have to express it by playing an instrument. I have tried to prime my auditory cortex experimenting with various types of noises, hoping that clear imagery experiences will somehow emerge, but without success. Oliver Sacks, who described himself as a verbalizer and poor imager, recalled how after taking amphetamine his weak imagery became very accurate and stable but his "abstract thinking was extremely compromised" (Sacks, 2007, p. 158).

I had hopes that playing an instrument may in the long run improve my musical imagery. Now, after more than 20 years of playing various electronic wind instruments almost every day, I still cannot imagine what I will play before actually playing it. I remember fingering patterns that produce notes of a particular pitch, so I can instruct myself to play some themes, like for example, A–C–E pattern. With a lot of repetitions I can remember some sequences of finger movements, but learning anything by heart is extremely tedious. It may look like amnesia for melodies, but they are somewhere in my long-term and procedural memories, although not accessible to sensory imagery. Improvising I cannot repeat what I have played a moment ago but may try to come back to consciously verbalized patterns. I have played some tunes many times but remember fragments of only a few. Each time I may play them slightly differently. With many years of experience, a lot of musical knowledge is internalized at a higher cognitive level. Covert imagery seems to contribute to my improvisations, despite the lack of inner auditory qualia. I may notice that I have made an error, but still can complete the phrase in a way that usually sounds right, improvising in a reasonable way[2].

The flexibility of my improvisation shows that it is not based on implicit learning only (Schacter, 2008). Experiments with implicit sequence learning have shown that learning higher-order sequential structures occurred only when distinct contexts were provided, while learning simple sequences could be done without context (D'Angelo et al, 2014). These experiments involved young typically developed university students. In this case, adding clear visual context can increase the flexibility of implicit sequence learning. In the case of a melody played on a wind instrument adding context will be more difficult. Perhaps the memory of the overall melody structure may provide some context. My experiments with playing along to MIDI tunes or to background tracks that can add more context to playing were not that successful. It is hard to play along when one is not able to anticipate what will be the next sound.

Playing music for someone with auditory agnosia is similar to maneuvering blindly in the auditory space, without knowing what will happen when the next action is taken. Each action is a step in the dark, without knowing the consequences. Various qualia of the motor, tactile, or emotional feeling types may be generated in this process, instead of the auditory qualia. The performance may be impaired, but improvisation sounds fine, and overall structure of the music that is played is maintained. This situation may be compared to the blindsight phenomenon, where an intuitive feeling based on some vague qualia guides the movements of a blind person (Weiskrantz, 1986). People capable of blindsight could see in the past, and even if their primary visual cortex is completely damaged alternative route from lateral geniculate nuclei to the higher visual areas is still working (Weiskrantz, 1996). They cannot experience visual qualia but may have some ability for visual orientation. By analogy in the case of total auditory imagery agnosia, the name "imagery deafness" could be used to capture this phenomenon.

Although I cannot achieve an expert level of music proficiency I have managed to give a few public concerts. Incidentally, I have never learned proper movements in more complex dances, despite some efforts already in adolescence. A good dancing ability seems to require imagery of movements. I am also not able to recall any images if I close my eyes. At rare times when I have dreams they contain some vague visual images, more like a feeling of knowing than a sensory experience. Aspirin and other drugs that increase the blood flow in the brain increase intensity of my dreams. Watching a movie before going to sleep activates memories of scenes, of what was happening but without visual or auditory elements. Remembering violent scenes seem to induce slight muscle contractions and proprioceptive feelings of movements, reminiscent of direct engagement.

Remembering words for songs that I find compelling is much easier than remembering poems or boring songs, perhaps because emotional arousal increases neuroplasticity. Recognizing instrumental melodies is also not a problem, but it is not based on melodic contours, rather a combination of rhythm/peak frequency patterns. Algorithms used to recognize music by Shazam and other popular software work on a similar principle. Probably the bottom-up stream sends sensory information correctly, enabling categorization of stimuli, but the top-down stream is too weak to re-create a vivid experience.

---

2  Examples of improvisations are here: https://www.is.umk.pl/~duch/prywatne/Fjukawka/index.htm

I gave several talks on imagery agnosia at conferences and met a few scientists who have no imagery in any sensory modality. This is also my personal experience. The shapes, colors, sounds, tastes, or smells do not appear in my imagination. Sometimes I experience something like a flash of understanding, a new idea, an insight that feels like a shadow of thought cast into the world of senses. My experience is quite similar to Watkins (2018), who called it "invisible imagery", pictures "too faint to see" but feels they are there. A few times in the hypnagogic state I've heard quite clearly a single word and assumed it must be unusually strong activation of my auditory cortex, something akin to hypnagogic jerk.

The obvious strategy to remember important stimuli is to verbalize them: remember some characteristic features of a person, the color of clothes, or type and color of a car. Learning mnemonic techniques based on imagery (methods of loci) does not work, but other techniques based on verbalization are useful (for example, π=3.1415926, as "How I wish I could recollect pi easily"). My autobiographical memory is also based on verbal recall of events rather than "mental time travel" to episodes in my life. I realized it only after reading Watkins' personal account (Watkins, 2018) and the paper on severely deficient and highly superior autobiographical memory (Palombo et al., 2018). I was surprised that people can vividly recollect personal experiences. Individual differences have been confirmed using neuroimaging and electrophysiological techniques (see the review in Palombo et al., 2018). Endel Tulving, who proposed the distinction between semantic and episodic memory in 1972 believed that animals do not have episodic memory, and some perfectly intelligent people also lack episodic memory. He thought that it is a matter of time before they will be found[3]. One way of working around autobiographical memory deficit is by making a lot of photographs that help to bring back autobiographical memories (I have now over 100 000 on my computer).

After over 40 years of marriage my wife, who is a strong visualizer, was shocked when I have clearly explained to her my imagery agnosia. It is fascinating that such differences in cognition go unnoticed for years.

**IMAGERY AGNOSIA**

Individual differences in cognition are surprisingly hard to notice. Imagery agnosia may be quite rare, but it is quite likely that it comes in many forms. Agnosia, the inability to extract relevant information from the bottom-up processing of sensory data, leads to specific cognitive deficits. Neuropsychology has identified at least 30 types of agnosia (Schoenberg & Scott, 2011). Calling some perceptual abilities a gift or treating it as agnosia depends on how the norm is defined, how common or rare some deficits are. For example, most people can estimate the relative pitch of notes played one after another. Comparing them to those rare people that are capable of perfect pitch and can name the pitch of individual notes, most people have strictly speaking pitch agnosia. In general, Caucasian population only 1 person in less than 10.000 has perfect pitch. However, among Caucasian music students in the USA it is about 9%, but for Chinese music students it is 65% (Deutsch et al. 2006). The deficit in normal pitch processing leads to amusia (congenital or acquired), affecting less than 4% of a general population (Peretz and Vuvan, 2017). This is true perceptual pitch agnosia, although hearing may otherwise be quite normal.

---

3   Cited in the web article: https://www.science.ca/scientists/scientistprofile.php?pID=20

The norm in the wider population decides what is called agnosia and what is a gift. In the case of imagery agnosia, we do not have good norms. Bottom-up processes responsible for many forms of agnosia may have their equivalents in top-down processes, responsible for imagery. Bottom-up processing of any aspect of sound perception may fail due to focal neurological deficits, leading to acquired amusia, problems with the hearing of pitch change directions, tone intervals, melodic contours and patterns, tonal structure, the timbre of instruments, and temporal structures such as intervals, rhythm, and meter, as well as musical memory and emotional responses (Stewart et al., 2006; Hubbard, 2019). Neural connections between primary sensory brain areas and higher sensory and associative areas are not symmetric, so specific imagery agnosia may not be correlated with true agnosia. Amusia and imagery agnosia are distinct phenomena.

This should also be true in relation to all sensory modalities. So far research trying to identify specific forms of imagery sensory agnosia has not been conducted. The division between total and voluntary aphantasia proposed by Watkins (2018) is very crude and not based on neural mechanisms. A new branch of neuropsychology devoted to such studies is needed (Duch, 2009; 2011; 2012). Imagery agnosia has been ignored for a long time because it is not correlated with obvious behavioral impairments. Faw (2009) estimated the percentage of the general population that have no visual imagery at 2-5% and the complete lack of auditory imagery in only 2% of people. Surprisingly about 12% of his respondents could not internally hear a song but could hum it. This is an indication that the strength of coupling between memory for sequences with motor and sensory areas may be quite different, and we may be able to express our "hidden imagery" via motor route, but not the internal route that activates sensory cortices.

Zeman (2016) has described 21 people with total visual imagery agnosia, calling this phenomenon "aphantasia". At the other extreme, there are people with "hyperphantasia". Extreme kinds of imagery appear to be relatively rare, but on a global scale thousands of people have been found through the Internet with such conditions, creating great interest in imagery research (Zeman et al., 2020). Although the name "aphantasia" is attractive this phenomenon should be properly classified in neuropsychology as a particular form of imagery agnosia. Visual agnosia is the inability to process some aspects of visual information. Total visual agnosia is called blindness, but agnosia may also be quite specific: akinetopsia is the inability to perceive visual motion, achromatopsia to perceive colors, associative visual agnosia to recognize specific category of objects, apperceptive visual agnosia to form complete percept. Recognition of complex visual stimuli may lead to prosopagnosia, or "face blindness", but it may also impair recognition of animals. Pure alexia is the inability to recognize text. Agnosia may touch all aspects of sensory perception, basic that are normally extracted at the primary sensory cortex level, or complex perception that requires involvement of tertiary sensory areas integrating information about different aspects of sensory experience. We should suspect that all forms of agnosia have their imagery counterparts. Perhaps some people who have visual agnosia are not able to imagin faces or letters. Research of imagery agnosias has barely scratch the surface. Using such name as "aphantasia" is obscuring the horizon.

Recent research based on questionnaire data from 2000 participants (Zeman et al., 2020) showed that imagery agnosia (aphantasia) is associated with occupations requiring abstract thinking, such as computer and mathematical sciences (Watkins, 2018), while hyperphantasia or extremely vivid imagery is more frequent in arts, media, and entertainment. Most people with imagery agnosia reported some visual experiences in dreams. This indicates that visual experiences do not require the involvement of primary sensory cortices, in agreement with the well known fact that people who lost their sight have visual dreams (Watkins, 2018; Sacks, 2003). Prosopagnosia is also associated with a lack of visual imagery. General memory tests do not show correlations with the vividness of imagery. Surprisingly, despite the lack of sensory imagery tests in tasks requiring imagination are performed at a high level of competence. It seems that imagery can be concealed and influence behavior without conscious experience (Schwitzgebel, 2011).

Objective measures of extreme imagery based on fMRI neuroimaging in the resting state (Milton et al., 2021) show stronger prefrontal and visual network connectivity in people with hyperimagery. Event-Related Brain Potentials (ERPs) may also be well suited here: after several presentations of simple melodies removing one of the notes should show ERPs reaction for missing sound if the person has vivid imagery (Alain & Winkler, 2012; Janata, 2001). The anticipation of the missing sounds is sufficient to create ERP response, correlated with the vividness of the auditory imagery questionnaire, as they are in the case of vision. In people with imagery agnosia, auditory ERP response is not expected. The only way that someone without auditory imagery may know that there is a melody in her/his head is by humming or playing it on an instrument. What kind of brain response should be expected? The imagery may elicit various activations, but not in the sensory cortices. It would be interesting to compare ERP from central and temporal electrodes.

Neuroimaging/lesion studies show that damage to the auditory areas along the superior temporal gyrus impairs pitch discrimination and melodic contour processing (Hubbard, 2019). In many cases, there is no obvious change in neural function that current neuroimaging tools could discover. Amusia may be due to the wrong wiring between auditory cortex areas, or to the inability of neurons in the auditory cortex to precisely synchronize their activity. Cognitive models of music processing do not include imagery and top-down processes (Peretz et al., 2003; Peretz and Vuvan, 2017).

Perception, memory, and imagery share some brain areas but may be treated as three partially independent dimensions (Dijkstra et al. 2017; Winlove et al. 2018). Mental imagery is also related to emotions (Holmes & Mathews, 2010). Pleasure related to music and art is the fourth dimension. People may enjoy music despite poor musical memory, imagery, or even congenital amusia (McDonald & Stewart, 2008). Large scale tests of 20 000 participants showed that only 1.5% have true amusia (Peretz and Vuvan, 2017), manifesting itself in poor pitch discrimination (tone deafness). Congenital amusia is hereditary, with 46% first-degree relatives similarly affected and 35% unsure whether this is the case. Musical training is not effective in such cases. Associations between congenital amusia and spatial orientation problems have been found in 15% of cases.

Almost half of the congenital amusics listen and appreciate music, so there is clear evidence for a dissociation between impaired music perception and music appreciation. "The psychology of talent" book (Edenborough and Edenborough, 2011) provides many definitions of talent, focusing on special natural capacity for excellent, almost perfect performance. A key aspect of musical talent is the ability to imagine music. Musicians need to connect the impression they desire to create with a "feel" they know will produce particular sounds. The mental image is a memory reference that enables the selection and initiation of motor actions leading to the creation of desired sounds. Is the vivid internal auditory image really necessary to play music? Must the anticipatory processes be fully conscious, leading to the experience of all qualia before the actual sound is produced, or can they be hidden from sensory experience, controlling motor actions? Does it make sense to talk about hidden or unconscious imagery?

The emotional component is based on yet another mechanism of information flow in the brain. People who love music have a strong coupling between the association cortex and positive valence subcortical and cortical structures, deriving pleasure from listening to music. It would be very surprising to find good musicians with auditory imagery agnosia. Tests for musical imagery should be taken into account in art and music education.

Reviewing mental imagery Pearson et al. (2015) write "Recent research suggests that visual mental imagery functions as if it were a weak form of perception." They also suppose that "mental imagery is conceived of as a type of top-down perception". Neuroimaging studies led Borst and Kosslyn (2018) to the conclusion that mental imagery and perception are perceived in a similar way. The view of imagery as a kind of simulation of perception is now widely accepted in cognitive sciences (Thomas, 2021). Activity patterns in sensory cortices encode perceptual and mental images, but re-activation of such images requires strong top-down neural projections. This process may be compromised in many ways leading to imagery agnosia.

### A bit of neurobiology

What biological factors may be responsible for imagery agnosia? Normal perception requires top-down influences to form expectations and help to establish conscious percepts (Gilbert & Sigman, 2007). Many neurobiological mechanisms may cause developmental differences in density and properties of different types of neurons, influencing the excitability of neurons and the strengths of their connections. Genetics and environmental factors are responsible for the quality of sensory transduction and discrimination of sensory signals, shapes, colors, movements, pitch and chroma, timbre, rhythm, melodic contours, harmony, and many other aspects of visual and auditory experiences. Extraction of such features from information streams in the brain may fail for many reasons, resulting in various types of agnosia.

In the book "The Living Brain" (W.G. Walter, 1953) noted that the EEG alpha oscilations are strong in the eyes shut condition, and disappear when eyes are opened. In people that reported poor visual imagery amplitude of alpha rhythms was not dependent on opening or closing eyes. Top-down projections in different pathways that reach primary sensory cortices may not be sufficiently strong to elicit robust activation that will lead to vivid imagery and desynchronization of alpha rhythm. Desynchronization may result from an increase in the irregularity of the EEG signal (Stam, Tavy and Keunen, 1993). Environmental factors, education, and culture induce neuroplastic changes changing brain connections. For example, musical ear training usually improves pitch discrimination. The second major factor depends on the internal properties of neurons. Keogh, Bergmann, and Pearson (2020) using brain imaging and transcranial magnetic stimulation to decrease visual cortex excitability and induce the experience of phosphenes showed that the strength of sensory imagery is predicted by lower neural resting activity and excitability levels.

The hypothesis proposed here is that individual differences are based on the density and strength of the top-down and bottom-up projections between primary sensory cortices and associative cortex responsible for memory and imagery, and projections to limbic structures involved in positive and negative valence. This may contribute to different types of talents, with strong imagery typical for artists and musicians, and weak imagery in the case of logicians, mathematicians, professional chess players, and other people who work on abstract, conceptual problems. Experiments by Blazhenkova and Kozhevnikov (2016) suggest that object visualization (ventral visual stream, involving temporal and occipital cortex) is related to artistic creativity, while visualization of spatial relations (dorsal visual stream, involving parietal cortex) is related to scientific creativity. This observation supports the parietal-frontal theory of intelligence (Colom et al. 2010) and points to a higher probability of imagery agnosia prevalence in creative people working in abstract domains. It also agrees with the simplified model of the experiential learning styles based on information flow in the brain (Duch, 2020). Activation of sensory cortices during abstract reasoning may not contribute to efficient problem-solving and may be a waste of brain energy. Abstract thinking is facilitated by imagery without sensory qualia with weak top-down brain activations that are not sufficient to invoke visual images or auditory sensations (see the personal account of Watkins, 2018). Oliver Sacks has described higher-order mental imagery in a very compelling way: sensory imagination is "utterly different from the higher and more personal powers of the imagination […]. It is by such imagination, such "vision," that we create or construct our individual worlds." (Sacks, 2003, p. 59). This kind of thinking style contributes to the efficient use of energy by the brain in such fields as algebra, logic, or theoretical physics.

Rough relations between activations of brain systems involved in the perception of music and art are presented in the figure below. Primary and higher visual and auditory sensory cortices receive strong inputs from associative cortices where long-term memory is stored. Imagery is a result of working memory activation that draws on long-term memory. Activation of the cortex influences the limbic structures such as the nucleus accumbens (Nacc) and the amygdala, adding negative and positive valence to brain states induced by sensory experience in a specific memory-based context.

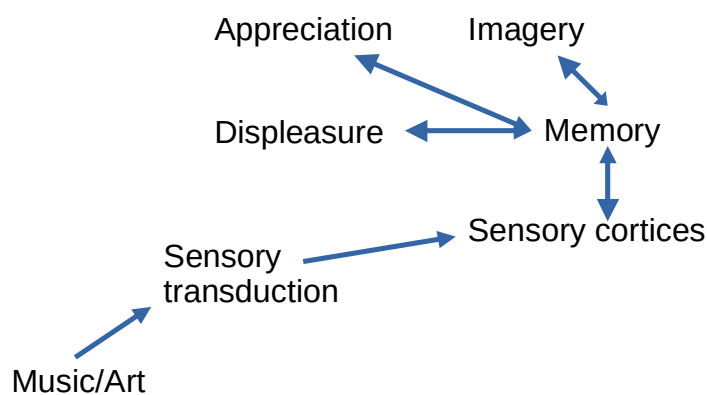

*Fig. 1: Rough relations between activations of brain systems involved in the perception of music and art.*

**Prevalence and individual differences**

Distribution of the vividness of imagery has non-Gaussian distribution, with extreme cases of very vivid imagery much more frequent – about 30% of respondents – than the lack of it – less than 5% of respondents (Faw 2009). Is there a common biological factor behind imagery in different modalities – visual, auditory, tactile, olfactory, gustatory, kinaesthetic, and bodily sensations? Correlations between smell and taste, or touch and kinaesthetic imagery are high, while between visual and smell imagery are quite weak (Faw 2009). Subjects may rate one modality as strong and the other as weak. In the case of vision and auditory imagery tests using The Plymouth Sensory Imagery Questionnaire (Andrade et al. 2014) showed that 10% of participants had scores in the top third on vision but the bottom third on sound, 8% the bottom third on vision, and the top third on sound. Only 13% got high scores in both modalities, and the same percentage got both scores in the lower third. In the Majjid et al. (2018) study significant differences in the hierarchy of importance of sensory experiences among people from 20 different cultures were found.

Individual differences in average imagery scores may be stronger due to environmental factors than genetic influences. However, in the case of imagery agnosia, genetic factors seem to have a much stronger influence than in average vividness of imagery (Zeman et al., 2020). Analysis of 2000 cases of people with imagery agnosia (aphantasia) showed that 54% do not experience imagery in any sensory modality, and about 1/3 had at least one other sensory modality unaffected (in normal or vivid range). The correlation of vividness of imagery within the family also points to the importance of genetic factors. In the largest controlled screening study prevalence of absent or dim/vague imagery measured using the VVIQ test was slightly lower than 4% while total absence of imagery had prevalence of 0.8%.

**SUMMARY**

I have described here phenomenology of imagery agnosia, with focused on auditory imagery agnosia, based on personal experiences. After more than 20 years of experimentation with learning to play music, my sensory imagery did not emerge, although my higher-level covert imagery has certainly improved. We learn about the world interpreting sensory experiences, and in most cases can partially recreate such experiences thanks to episodic memory. These experiences can be decomposed into a large number of components, features derived from sensory signals by specific brain mechanisms. Over 30 types of agnosia (Schoenberg & Scott, 2011), such as amusia or achromatopsia, and even more complex visuospatial deficits, such as prosopagnosia, show in how many ways these mechanisms may fail. They are based on bottom-up information processing, leading to the formation of episodic memories and activation of recognition memory.

The imagery of visual, auditory, or other sensory information needs to re-create conscious qualia in proper modalities by reactivating sensory cortices in a top-down manner (Borst & Kosslyn, 2008). Among many factors that contribute to vivid imagery, the strength of top-down connections (Gilbert & Sigman, 2007) and the excitability of sensory cortices (Keogh, Bergmann & Pearson, 2020) are probably the most important. These two factors may explain individual differences in imagery vividness and autobiographical memories. Depending on the strength of connectomes and activation flow through different brain pathways they should elucidate various forms of imagery agnosia or its opposite, extremely vivid imagery. All forms of agnosia may have their analogs in imagery agnosia. So far research has focused on visual imagery, calling it "aphantasia", but this is a vast subject that should be properly placed among other phenomena belonging to neuropsychology. For that reason I proposed to call it "imagery agnosia".

Research on imagery agnosia is relevant not only to neuropsychology, but also to the phenomenology of conscious experiences, metacognition, autobiographical memory, psychology of individual differences, understanding artistic and musical talent, aesthetic experience, learning different skills, early education, or influence on mathematical abilities in STEM professions. Many interesting questions in this area remain unanswered.

**Acknowledgment**:
Supported by the National Science Center grant UMO-2016/20/W/NZ4/00354.
This work is dedicated to Piotr Francuz, with whom we have discussed relations between imagery and talent several times.